\documentclass[twocolumn,10pt,a4paper]{article}
\usepackage[utf8]{inputenc}
\usepackage[margin=1.9cm]{geometry}
\usepackage{graphicx}
\usepackage{amsmath}
\usepackage{amssymb}
\usepackage{amsfonts}
\usepackage{mathtools}
\usepackage{physics}
\usepackage{xcolor}
\usepackage[colorlinks=true,allcolors=blue]{hyperref}
\usepackage{authblk}
\usepackage{caption}
\usepackage{abstract}
\usepackage{titlesec}
\usepackage{subcaption}
\usepackage[T1]{fontenc}
\titlespacing*{\section}{0pt}{1.1ex plus .3ex}{0.6ex}
\titlespacing*{\subsection}{0pt}{0.9ex plus .2ex}{0.4ex}
\titleformat{\section}{\normalfont\large\bfseries}{\thesection}{0.6em}{}
\titleformat{\subsection}{\normalfont\normalsize\bfseries}{\thesubsection}{0.5em}{}
\newcommand{\Parties}{\mathcal{P}}
\newcommand{\bigDelta}{\mathop{\Delta}\limits}

\providecommand{\keywords}[1]
{
  \small	
  \textbf{\textit{Keywords:}} #1
}

\newtheorem{theorem}{Theorem}
\newtheorem{lemma}{Lemma}
\newtheorem{corollary}{Corollary}

\title{Efficient Multiparty Entanglement Distribution in Dynamic Quantum Networks}

\author[1]{Roberto Negrin}
\author[2]{Nicolas Dirnegger}
\author[1]{William Munizzi\textsuperscript{\normalfont\textasteriskcentered}}
\author[1]{Jugal Talukdar}
\author[1,2]{Prineha Narang\textsuperscript{\normalfont\textdagger}}

\affil[1]{\small Division of Physical Sciences, College of Letters and Science, University of California, Los Angeles}
\affil[2]{\small Electrical and Computer Engineering Department, University of California, Los Angeles}

\affil[ ]{\small \normalfont\textasteriskcentered~munizziwilliam@gmail.com \\ \normalfont\textdagger~prineha@ucla.edu}
\date{}

\begin{document}

\twocolumn[
\maketitle
\begin{onecolabstract}
Distributing multipartite entanglement over a quantum network means routing it through a shared resource state, and existing measurement-based schemes do so by searching for a fresh path and re-verifying the current topology before every request. In a dynamic network, where links fail and recover between requests, this step is paid again on every request and depends on up-to-date global topology, placing a network-wide classical exchange on the critical path of each entanglement request. We introduce DODAG-X, which removes it. A single Destination-Oriented Directed Acyclic Graph (DODAG) spanning tree is computed once and reused across all requests, so each party's route is recovered by following parent pointers instead of a new search. The per-request routing cost drops from $\mathcal{O}(N)$ to $\mathcal{O}(\sqrt{N})$ on symmetric grids and $\mathcal{O}(\log N)$ on small-world networks, and only the $N-1$ tree links need be maintained under link loss. Routing on the sparse tree also shrinks the neighborhoods that must be cleared to isolate the parties, lowering measurements per request by ${\approx}19\%$ on small-world graphs and up to $34\%$ on moderately dense, strongly rewired ones, with the savings concentrated in the isolating step; on a fixed tree the two protocols use identical counts. We prove correctness for up to three parties with no restriction on topology, and prove a sufficient condition under which one application yields an $n$-party GHZ state for any $n$; we then delimit it, exhibiting requests outside the hypothesis whose output is genuinely multipartite entangled yet in a different local-Clifford class. Under a discrete-time Markov failure model the classical repair layer matches the reachability of full-graph re-search up to a failed-edge fraction $f\approx0.5$, and a coherence criterion $T_2\gg d\cdot\tau_c$ identifies the hardware platforms on which the scheme is viable.
\end{onecolabstract}
\vspace{1em}]


\section{Introduction}

Quantum networks underpin distributed quantum computation and quantum key distribution \cite{thalacker2021anonymous,belghachi2023quantum}, but are limited by noise and loss in long-distance entanglement generation \cite{wengerowsky2019entanglement,kaushal2017free}. While two-party entanglement distribution is well studied \cite{pirandola2019end,Abane:2024ipy}, efficient multipartite distribution remains open. A standard approach models the network with graph states, which are multipartite entangled states in which each vertex is a qubit and each edge a correlation \cite{hein2006entanglement}. Graph states are used across measurement-based computing \cite{li2023robust}, error correction \cite{hilaire2021error}, secret sharing \cite{cirac1999distributed}, and metrology \cite{shettell2020graph}.

The X-protocol \cite{hahn2019quantum} entangles two (or up to four) nodes using only local measurements, without first isolating the subsystem. It was later extended toward $n$ parties under a vertex-minor condition \cite{mannalath2023multiparty}. Its drawback is a path-finding step before each entangling operation: the search is repeated for every request, requires up-to-date global state, and optimal path-finding is hard in general. 

Classical networking faces a structurally similar problem. Low-Power and Lossy Networks (LLNs) consist of unreliable nodes communicating over intermittent links, where the energy and messaging budget of the devices rules out on-demand route discovery. The standard solution, the RPL routing protocol \cite{rfc6550,vasseur2011rpl}, organizes nodes into a Destination-Oriented Directed Acyclic Graph (DODAG): a spanning structure rooted at a single destination in which every node holds only a parent pointer and its own rank, so a route is followed instead of searched for. When a link fails, the orphaned node reattaches through a surviving neighbor by local classical messaging, and the rest of the structure is untouched. Routing therefore survives an unreliable substrate without any node holding a global view.

For quantum networks, DODAGs are attractive: maintaining a tree means holding $N-1$ entangled links in place of the full edge set, and the maintenance is purely classical, adding no quantum overhead \cite{vasseur2011rpl} beyond re-establishing the lost links themselves. Ref.~\cite{yang2023asynchronous} takes this step for pairwise entanglement, using a DODAG to maintain a two-party entanglement structure under link loss at higher entangling rates than synchronized swapping. Multipartite routing on such a structure is missing, and DODAG-X supplies it: a measurement-based routing procedure run directly on the DODAG, inheriting both the amortized path-finding and the classical repair layer. Other works optimize graph-state generation directly \cite{Sen2024,Fan2024,Meignant2019}; our target is the routing of entanglement over a resource state already in place.

We combine the measurement-based routing of the X-protocol with a semi-classical DODAG that preserves a tree entanglement structure under link loss, with maintenance handled by the preconstructed DODAG layer of \cite{yang2023asynchronous}. We prove correctness for up to three parties in any network, and prove a sufficient condition for single-shot GHZ distribution at any party number (Theorem~\ref{thm:ghzn}). Throughout, we distinguish the physical network, with qubits as nodes and channels as edges, from the entanglement structure or instant topology \cite{yang2023asynchronous} that routing manipulates; Sec.~\ref{sec:dodag} develops the distinction. On a fixed tree DODAG-X uses exactly as many measurements as the X-protocol (Theorem~\ref{thm:measurement}), so the reported savings have two other origins. The first is structural: routing on a sparse spanning tree embedded in the physical network yields smaller neighborhoods to clear with isolating $Z$-measurements, which lowers the total count relative to running the X-protocol on the denser physical graph. The second is classical: committing to one precomputed tree replaces repeated per-request path search and graph re-verification with a one-time setup.

\begin{figure*}[t]
    \centering
    \includegraphics[width=\linewidth]{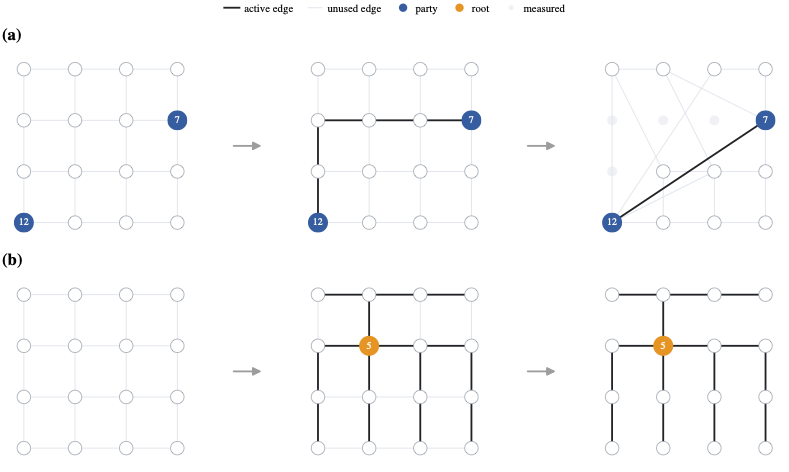}
    \caption{Routing on the physical network versus on an embedded tree, shown
    on the same $16$-node grid. Active edges are black, unused edges pale gray;
    parties are blue and the DODAG root is orange.
    \textbf{(a)}~The X-protocol. Two parties are selected, the shortest path
    between them is computed on the physical graph, and each interior node
    along it is $X$-measured using one party for local complementation. The
    rerouted correlations leave the two parties directly connected, but
    embedded in a denser residual graph (right, with measured vertices shown as
    faint disks), so a further round of $Z$-measurements on their neighborhoods
    is required to isolate the pair in an EPR state. This path search, together
    with a re-verification of the current topology, recurs for every request.
    \textbf{(b)}~The DODAG construction on the same lattice. A spanning tree
    rooted at a node of minimum eccentricity is built once, classically, from
    the physical network; only its $N-1$ links must be maintained under link
    loss. The resulting entanglement tree (right) is the structure on which
    DODAG-X routes, giving every node a unique parent and a unique path to the
    root.}
    \label{fig:topologies}
\end{figure*}

We treat link losses as edge removal and verify, under a dynamic failure-and-recovery model, that the classical repair layer keeps the tree routable at success rates comparable to per-request re-search (Sec.~\ref{sec:dynamic}); we do not model decoherence or compute fidelities. We assume the network is initialized in the graph state $\ket{T}$ of the DODAG tree, established by the same periodically repeated entanglement-generation step that the maintenance layer of \cite{yang2023asynchronous} uses to restore lost links (Sec.~\ref{sec:dodag}), at a one-time cost of $N-1$ entangling operations. Where a platform natively prepares the full physical graph state $\ket{G}$, the tree state is recovered by applying $CZ_{u,v}$ along each non-tree edge, since $CZ$ is its own inverse; this costs $|E|-(N-1)$ two-qubit operations along existing physical channels, paid once per network rather than per request.

\section{Background}\label{sec:background}

\subsection{Graphs and graph states}\label{sec:graphs}

A graph $G=(V,E)$ consists of vertices $V$ and edges $E\subseteq V\times V$. We take all graphs simple and undirected, so an edge is an unordered pair $\{u,v\}$ with $u\neq v$ and $\{u,v\}=\{v,u\}$. A path $P_{0,k}=\{v_0,\dots,v_k\}$ is a sequence of distinct vertices with $\{v_i,v_{i+1}\}\in E$ for all $i$. The neighborhood of $v$ is
\begin{equation}\label{eq:neighborhood}
    N_v=\{u\in V \mid \{u,v\}\in E\},
\end{equation}
which never contains $v$ itself. Deleting a vertex removes it together with every
incident edge,
\begin{equation}\label{eq:delete}
    G-v = G\big(V\setminus\{v\},\ \{\{i,j\}\in E: i,j\neq v\}\big),
\end{equation}
while deleting an edge leaves the vertex set intact. The former corresponds to a
measurement on the physical system (Eq.~\eqref{eq:z}); the latter corresponds to the
loss of a communication channel, which is how we model link failure in
Sec.~\ref{sec:dynamic}.

\emph{Local complementation} at $v$ replaces the subgraph induced on $N_v$ by its
complement: edges among neighbors of $v$ are deleted, and non-edges among them are
created, with all other edges untouched. Formally,
\begin{equation}\label{eq:lc}
    \tau_v(G) = \big(V,\ E\,\Delta\,\{\{i,j\}\mid i,j\in N_v,\ i\neq j\}\big),
\end{equation}
with $\Delta$ the symmetric difference \cite{ehrenfeucht2004transitivity}. Every
measurement-driven step of the protocols below reduces to local complementation
together with vertex deletion.

In a quantum network, vertices (``nodes'') are qubits or registers of qubits and
edges are communication channels. The corresponding quantum object is a
\emph{graph state} $\ket{G}$, in which vertices are qubits and edges are two-qubit
correlations \cite{hein2006entanglement}:
\begin{equation}\label{eq:graphstate}
\ket{G} := \bigotimes_{\{u,v\}\in E} CZ_{u,v}\,\ket{+}^{\otimes|V|},
\end{equation}
with each qubit initialized in $\ket{+}=(\ket{0}+\ket{1})/\sqrt{2}$ and
entanglement generated by controlled-$Z$ gates. Eq.~\eqref{eq:graphstate} defines
the state; other preparation routes are available
\cite{kaur2024resource,kay2006graph}. Two graph states are LU-equivalent if they
agree up to a local unitary, and local Clifford operations always map $\ket{G}$ to
an LU-equivalent state \cite{hein2006entanglement,van2004graphical}.

Local Pauli measurements act directly on the underlying graph
\cite{hahn2019quantum}. A $Z$-measurement on $v$ deletes it,
\begin{equation}\label{eq:z}
    Z_v(G) = G - v,
\end{equation}
leaving correlations among $V\setminus\{v\}$ intact up to a local Pauli. An
$X$-measurement gives
\begin{equation}\label{eq:x}
    X_v(G) = \tau_w Z_v \tau_v \tau_w(G),
\end{equation}
for a chosen $w\in N_v$. Eq.~\eqref{eq:x} is a $Z$-measurement conjugated by local
complementations: before $v$ is deleted, its correlations are rerouted onto the
neighbor $w$. A sequence of $X$-measurements therefore carries entanglement along
a path, and both the X-protocol and DODAG-X are built on this. We write $Z[v]$
and $X[v,w]$, and leave the outcome-dependent local Clifford byproducts implicit,
since different choices of $w$ yield graphs related by local unitaries
\cite{hein2004multiparty,dahlberg2018transforming}.

\paragraph*{Notation.} Two neighborhood notations recur. $N_v^{(m)}$ denotes the
neighborhood of $v$ after $m$ measurements have been applied (superscript $=$ a
count), while $N_v^{(u)}$ denotes the neighborhood of $v$ after measuring the
specific node $u$ (superscript $=$ a vertex label). We state which is meant
wherever either reading is possible.

\subsection{The X-protocol}\label{sec:xprotocol}

The X-protocol \cite{hahn2019quantum} entangles two nodes $i,j$ using only local
measurements, without first isolating the subsystem:
\begin{enumerate}\itemsep2pt
    \item Find the shortest $i$--$j$ path, e.g.\ with BFS or Dijkstra.
    \item $X$-measure each interior node of that path, excluding $i$ and $j$.
    \item $Z$-measure every node in $N_i\cup N_j$, again excluding $i$ and $j$.
\end{enumerate}
The result is an EPR pair on $i,j$; the three-party case is analogous \cite{hahn2019quantum}. The generalization to $n$ parties \cite{mannalath2023multiparty} requires the network to contain the target graph as a vertex minor, so that successive local complementations and vertex deletions distill it; on a repeater line this amounts to an additional node inserted between each pair of consecutive intermediate nodes along the path. Neither \cite{hahn2019quantum} nor \cite{mannalath2023multiparty} assumes a specific topology.

A single shortest-path computation is inexpensive ($\mathcal{O}(|E|+|V|\log|V|)$ with Dijkstra \cite{ashish2021path}), but it must be re-run for every request, together with a re-verification of the current topology, and it requires up-to-date global network state. DODAG-X amortizes this recurring cost.

\subsection{Dynamic networks and the DODAG}\label{sec:dodag}

In a static network all channels are operational at all times, which permits classical offline pre-processing. A realistic quantum network is instead dynamic: channels fail and recover over time through loss, noise, and decoherence \cite{kozlowski2020designing,michail2018elements}. We therefore distinguish the \emph{physical topology}, the hardware channels (fiber or free-space links) between qubits, from the \emph{virtual} or \emph{instant topology} \cite{yang2023asynchronous}, the entanglement structure among nodes, which need not follow any direct hardware connection (Fig.~\ref{fig:topologies}(b)). Entanglement distribution is the manipulation of physical nodes to realize a desired virtual topology. A routing protocol for such a network must find paths on demand and re-route when a previous path is lost, using only local information: each node knows the status of its incident links and its parent pointer, assigned during a one-time classical setup, and no global topology knowledge is required per request. Broadcasting entanglement status globally is slow and exposes the register to decoherence while it waits.

A Destination-Oriented Directed Acyclic Graph (DODAG) is a spanning tree rooted at a destination node that preserves the virtual topology under link loss, reconnecting orphaned nodes by classical messaging \cite{yang2023asynchronous,vasseur2011rpl}. DODAGs are standard in classical Low-Power and Lossy Networks, where the RPL protocol uses them to organize nodes hierarchically about a root and to route efficiently under intermittent connectivity \cite{rfc6550,vasseur2011rpl}. Maintaining the tree is a purely classical process and carries no quantum overhead \cite{vasseur2011rpl}, apart from re-establishing lost entangled links. For this to be physically viable the memory coherence time $T_2$ must exceed the classical latency of the repair messaging; we bound this regime in Sec.~\ref{sec:discussion}. Under that assumption the tree is logically static for the duration of one request.

We use a non-greedy DODAG \cite{rfc6550} and assume that a prior, periodically repeated, step has established an entangled tree across all nodes (Fig.~\ref{fig:topologies}(b)). Committing to a tree also means maintaining only $N-1$ entangled links instead of the full edge set, which lowers the standing memory and maintenance burden independently of the routing and measurement savings below \cite{dahlberg2019link,khatri2019practical}.

\begin{figure*}[t]
\centering
\includegraphics[width=\linewidth]{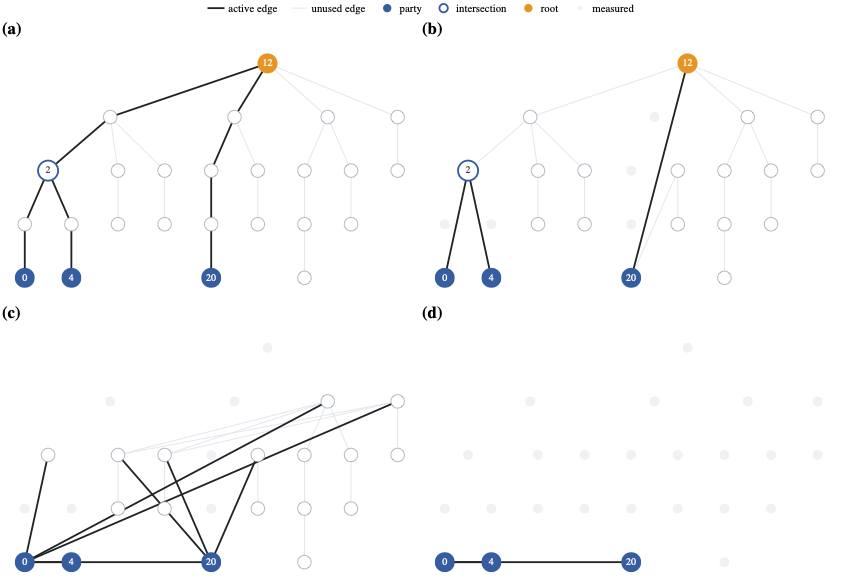}
\caption{DODAG-X on a $5\times5$ grid, case 5(c), drawn in the tree layout of the DODAG rather than in lattice coordinates. The DODAG is rooted at $r=v_{12}$ (orange), a node of minimum eccentricity; the parties $v_0$, $v_4$, $v_{20}$ are blue and the intersection node is the open node with a blue ring. Active edges are black, unused edges pale gray, and faint disks mark vertices already consumed by measurement. All four panels share one fixed frame and one set of node coordinates, so vertices can be tracked as the network is measured away. \textbf{(a)}~The unique path $P_{i,r}$ from each party to the root (Step~\ref{p:2}) and the intersection node $a_{0,4}=v_2$ at which the paths from $v_0$ and $v_4$ first meet (Step~\ref{p:3}); $v_{20}$ meets them only at the root. Since $a_{0,4}$ is neither the root nor a party, this is case 5(c). \textbf{(b)}~$X$-measurements along each party-to-intersection path entangle $v_0$ and $v_4$ with $a_{0,4}$ and $v_{20}$ with the root (Step~\ref{p:4}). \textbf{(c)}~The intersection node is entangled with the root along $P_{a_{0,4},r}$, and a final $X$-measurement at $r$ brings all three parties into a common entangled state (Steps~\ref{p:5}--\ref{p:6}); the parties are connected but still embedded in a dense residual neighborhood. \textbf{(d)}~$Z$-measurements remove every remaining neighbor of a party (Step~\ref{p:7}), leaving the isolated triplet as a three-vertex path centered on $v_4$, locally equivalent to a GHZ state.}
\label{fig:example}
\end{figure*}

\paragraph*{Root selection and construction.} The tree is built once, classically,
before any measurement. We select a root of minimum eccentricity
(Appendix~\ref{app:complexity}) and grow a spanning tree by breadth-first search,
giving every node a unique parent and a unique path to the root. A shallow tree
shortens party-to-root paths, which set the measurement cost. Construction costs
$\mathcal{O}(|V|+|E|)$; exact minimum-eccentricity root selection costs
$\mathcal{O}(|V||E|)$ and may be replaced by approximate-center heuristics at
scale. This setup is paid once per network and is distinct from the per-request
cost of Sec.~\ref{sec:complexity}. During a request the root may be reassigned to
the node on all party-to-root paths farthest from the current root
(Step~\ref{p:2}); on a tree this only relabels an existing node and leaves edges,
parents, and path lengths unchanged.

\section{The DODAG-X protocol}\label{sec:protocol}

We entangle a set of up to three parties $\Parties = \{i, j, k\}$ in an $N$-node network with a chosen DODAG. The protocol is illustrated in Fig.~\ref{fig:example}.
\begin{enumerate}\itemsep2pt
\item \label{p:1} Select $n\le 3$ target parties $\Parties$.
\item \label{p:2} For each party $i \in \Parties$, find the path $P_{i,r}=\{v_1{=}i,\dots,v_m{=}r\}$ to root $r$, indexed by decreasing distance to $r$, so that $v_1=i$ and $v_m=r$. If nodes other than $r$ lie on $P_{i,r}$ for all $i$, set the one farthest from $r$ as the new root. This choice is unique: the vertices common to all party-to-root paths form a subpath of the tree ending at $r$.
\item \label{p:3} For parties $i\neq j$, the intersection is $I_{i,j}\equiv P_{i,r}\cap P_{j,r}$. The intersection node $a_{i,j}$ is its least element in the path ordering of Step~\ref{p:2}, that is, the element of $I_{i,j}$ farthest from $r$, with $a_{i,j}=a_{j,i}$.
\item \label{p:4} Entangle each party $i$ with its nearest intersection node. We identify $P_{i,a_{i,j}}$ (the shortest such path), then perform $X[v,i]$ on each interior $v\in P_{i,a_{i,j}}$ (excluding $i$ and $a_{i,j}$), using the party $i$ for local complementation; parties that are not themselves intersection nodes are routed first. After this step each party is connected to, or is, an intersection node (Theorem~\ref{thm:step1}).
\item \label{p:5} Entangle all intersection nodes with the root. There are three cases. In case (a) there is one intersection node, so $a_{i,j}=r$ and this step is done. In case (b) $a_{i,j}\neq r$ is itself a party, already connected to $r$ after Step~\ref{p:4}, and this step is done. In case (c) $a_{i,j}$ is neither root nor party. Along $P_{a_{i,j},r}=\{v_1{=}a_{i,j},\dots,v_m{=}r\}$ perform $X[v_s,v_{s-1}]$ for even $s$ (excluding $a_{i,j},r$). If $|P_{a_{i,j},r}|$ is even, add a final $X[v_{m-1},i]$ using a party $i$ entangled with $a_{i,j}$ after Step~\ref{p:4}. For $n\le3$ these three cases are exhaustive, since at most one intersection node can differ from the root; party sets with several distinct non-root intersection nodes arise only for $n\ge4$ and lie outside the specified step (Sec.~\ref{sect:higher_party}).
\item \label{p:6} If $r\in\Parties$ the protocol is complete. Otherwise perform $X[r,i]$ for any party $i$ already entangled with the root. Such a party always exists, without any further graph check: the only case in which no party is root-connected after Step~\ref{p:5} is case (c), and there the party $i$ used for local complementation in Step~\ref{p:5} is entangled with the root.
\item \label{p:7} $Z$-measure every node in the neighborhood of each party except the parties themselves (Appendix~\ref{app:counts}).
\end{enumerate}

The result is a maximally entangled $n$-party state ($n\le3$) disconnected from the rest of the network (Theorem~\ref{thm:step2}). Steps~\ref{p:1}--\ref{p:4} apply for any party number, and extending Steps~\ref{p:5}--\ref{p:7} beyond three parties is discussed in Sec.~\ref{sect:higher_party}.

\subsection{Per-request complexity}\label{sec:complexity}

Existing schemes re-verify the network and run a path search per request. This costs $\mathcal{O}(|V|+|E|)$ with BFS, which is $\mathcal{O}(|V|)$ per search on sparse graphs ($|E|=\mathcal{O}(|V|)$), or $\mathcal{O}(|E|+|V|\log|V|)$ with Dijkstra \cite{ashish2021path}, and it is paid again on each request. DODAG-X pays a one-time setup, after which each request recovers a party-to-root path by following parent pointers, at a cost set by tree depth and with no new search or re-verification.

For an $m\times\ell=N$ grid, the ratio $m/\ell$ sets the depth. For $m/\ell\approx1$ the per-request cost is $\mathcal{O}(\sqrt N)$ (Appendix~\ref{app:complexity}), a factor of $\mathcal{O}(\sqrt N)$ below a new $\mathcal{O}(N)$ search. For $m\gg\ell$ it approaches $\mathcal{O}(N)$ and the advantage diminishes. For Watts--Strogatz small-world graphs \cite{watts1998collective} (mean degree $k$, rewiring probability $p$), average distances scale as $\log N$, so the per-request cost is $\mathcal{O}(\log N)$, and that scaling applies to any network with small-world statistics.

\subsection{Higher party number}
\label{sect:higher_party}

\begin{figure}[t]
\centering
\includegraphics[width=\columnwidth]{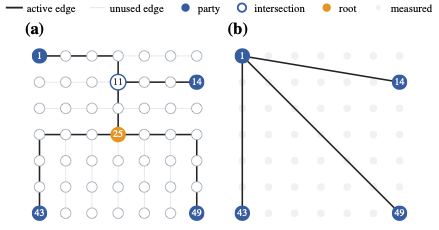}
\caption{A four-party request on a $7\times7$ grid, lying outside the
hypothesis of Theorem~\ref{thm:ghzn} yet still returning GHZ$_4$; conventions
as in Fig.~\ref{fig:example}. \textbf{(a)}~The DODAG rooted at $r=v_{25}$
(orange), with the parties $v_1$, $v_{14}$, $v_{43}$, $v_{49}$ (blue) and
their paths to the root. The parties $v_1$ and $v_{14}$ meet at
$a_{1,14}=v_{11}$ (open node with a blue ring), which is neither the root nor
a party, so Step~\ref{p:5} applies as case (c); every other pairwise
intersection node is the root. The pairwise intersection nodes therefore do
not all coincide, and the hypothesis of Theorem~\ref{thm:ghzn} does not hold.
\textbf{(b)}~The output after the root measurement of Step~\ref{p:6} and the
isolating $Z$-measurements of Step~\ref{p:7} is nonetheless the star
$K_{1,3}$ centered at $v_1$, the party used for local complementation in
Step~\ref{p:6}, locally equivalent to GHZ$_4$, in $27$ measurements total.}
\label{fig:higher}
\end{figure}

Steps~\ref{p:1}--\ref{p:4} generalize to $n$ parties: given a tree of size $N>n$, each party can be entangled with its nearest intersection node. When all pairwise intersection nodes coincide at a single vertex, equivalently when the parties occupy distinct branches of one vertex, which the re-rooting of Step~\ref{p:2} then makes the root, a single application of the protocol yields the star $K_{1,n-1}$, locally equivalent to the $n$-party GHZ state, for any $n$ (Theorem~\ref{thm:ghzn}). Appendix~\ref{app:higher} verifies the predicted star structure, including the location of its center, in both branches of the theorem, on four-party sets constructed to satisfy its hypothesis.

The hypothesis is sufficient but not necessary. Fig.~\ref{fig:higher} shows a four-party request whose paths define one non-root intersection node, so that Step~\ref{p:5} applies as case (c), and which still returns GHZ$_4$. The protocol as specified does not, however, handle several distinct non-root intersection nodes: Step~\ref{p:5} contracts a single one, and party sets producing more lie outside it. For $n\ge4$ a connected output no longer determines the entanglement class either, since connected graph states on four or more qubits split into several inequivalent LC classes \cite{hein2004multiparty}, only one of which is GHZ$_n$. Appendix~\ref{app:higher} exhibits a six-party request whose output is isolated and genuinely six-party entangled, yet LC-equivalent to the six-qubit linear cluster state instead of GHZ$_6$. When one application is insufficient, a concatenated construction in which the DODAG supports multiple qubits per node and the outputs of separate protocol runs are merged locally is one possible extension; we do not specify or cost it here.

\section{Performance benchmarks}\label{sec:benchmarks}

\begin{figure*}[t]
\centering
\includegraphics[width=\textwidth]{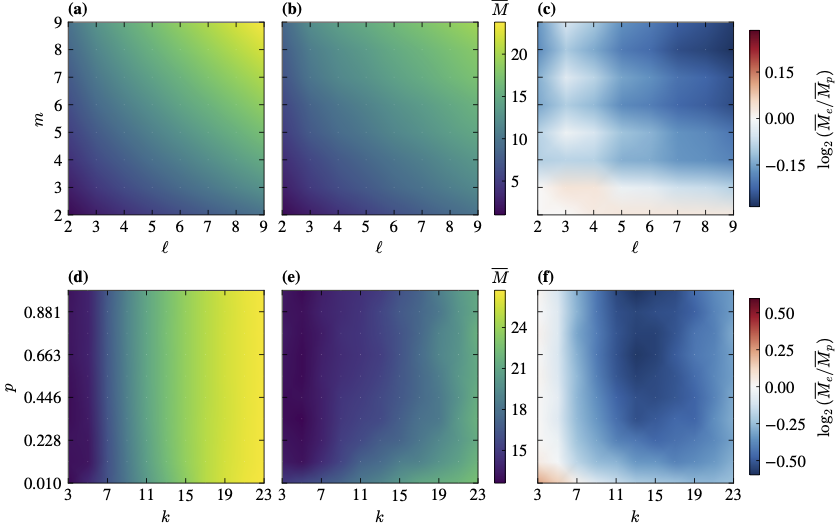}
\caption{Measurement-count benchmarks. Top row: grid lattices,
$2\le m,\ell\le9$, one deterministic instance per cell. Bottom row:
Watts--Strogatz small-world graphs ($N=30$), $40$ seeded instances per
$(k,p)$ cell. \textbf{(a)},\textbf{(d)}~Average measurements for the
X-protocol on the physical network; \textbf{(b)},\textbf{(e)}~DODAG-X on the
embedded tree, on a shared color scale with the panel to its left.
\textbf{(c)},\textbf{(f)}~Log-ratio $\log_2(\overline{M}_e/\overline{M}_p)$
[Eq.~\eqref{eq:logratio}]; blue where DODAG-X uses fewer measurements, red
where it uses more, white at parity. Of the $64$ grid cells, $53$ favor
DODAG-X, $8$ favor the X-protocol, and $3$ are at parity; of the $110$
small-world cells, $98$ favor DODAG-X and the $12$ regressing cells are
confined to the low-$k$, low-$p$ corner.}
\label{fig:benchmarks}
\end{figure*}

Following the measurement-based accounting of the X-protocol and its extension \cite{hahn2019quantum,mannalath2023multiparty}, we use the number of local measurements as our metric. Each measurement consumes part of the graph state irreversibly and is a potential error site, which makes the count a hardware-agnostic proxy for the cost of an entanglement request.

We compare the average number of measurements to entangle all triplets, using DODAG-X on the tree versus the X-protocol on the physical network. The percent reduction is defined as
\begin{equation}\label{eq:diff}
\frac{\overline{M}_p-\overline{M}_e}{\overline{M}_p}\times100,
\end{equation}
with $\overline{M}_p$ and $\overline{M}_e$ the average X-protocol and DODAG-X measurement counts, respectively. Grids are deterministic (one instance per $m\times\ell$ cell), and small-world cells are averaged over $40$ seeded instances with sampling standard errors. Difference maps plot the symmetric base-two log-ratio
\begin{equation}\label{eq:logratio}
\log_2\!\left(\overline{M}_e/\overline{M}_p\right),
\end{equation}
negative when DODAG-X uses fewer measurements. The two protocols consume different substrates: the X-protocol baseline routes on the full physical graph state, as specified in \cite{hahn2019quantum,mannalath2023multiparty}, while DODAG-X routes on the embedded tree. The comparison is between deployments, and this is deliberate: committing to the tree is the design choice under test, and its standing cost is lower, $N-1$ maintained links against $|E|$. Theorem~\ref{thm:measurement} separates the two contributions: on one fixed tree the protocols use identical measurement counts, so the reductions below are attributable to the substrate and not to the measurement schedule.

\subsection{Grid lattices}

We take $N=m\times\ell$ with $2\le m,\ell\le9$. The DODAG reduces neighborhood sizes, lowering the final $Z$-measurement count, while retaining path lengths comparable to the grid [Fig.~\ref{fig:benchmarks}(a,b)]. Of the $64$ cells, $53$ favor DODAG-X [blue in Fig.~\ref{fig:benchmarks}(c)], $8$ favor the X-protocol (red), and the remaining $3$ are at parity (white). Excluding the small-count edge cells $m,\ell\in\{2,3\}$, the mean reduction is $7.59\%\pm3.48\%$ (standard deviation across grid sizes, not sampling error). Including them it is $3.97\%$, since the smallest grids have such low measurement counts that a $\pm1$ difference produces large, sometimes negative, per-instance ratios. The largest single-cell reduction is $17.90\%$ at $m=8,\ell=9$ ($N=72$), and the red cells lie in this same small-count region.

The heatmap is not symmetric under $m\leftrightarrow\ell$ even though the grid graph is: minimum-eccentricity root selection and BFS construction are not orientation-invariant, so the off-diagonal asymmetry comes from the tree-construction heuristic. Performance therefore depends on the choice of spanning tree, and optimizing over trees is a possible improvement.

\subsection{Small-world networks}

We fix $N=30$ and scan $3\le k\le23$, $0.01\le p\le0.99$ [Fig.~\ref{fig:benchmarks}(d,e)]. DODAG-X improves most at high $k$, where local complementation modifies larger neighborhoods. There the X-protocol is roughly flat in $p$, while DODAG-X dips slightly at small $p$. Averaged over all cells, DODAG-X uses roughly $19\%$ fewer measurements ($19.44\%\pm11.24\%$ where the spread is the standard deviation across cells). The largest single-cell reduction is $33.85\%\pm0.81\%$ at $k=13,p=0.99$. The median per-cell sampling standard error is $0.70\%$, so the cell-to-cell spread reflects topology, not sampling noise. In the low-$k$, low-$p$ corner the DODAG can be unfavorable: the worst cell, $k=3,p=0.01$, uses about $18.6\%$ more. These $12$ regressing cells are confined to that corner, and DODAG-X is favorable across the other $98$ of $110$ cells [Fig.~\ref{fig:benchmarks}(f)].

To isolate the effect of $p$, we fix $k=9$ and vary $p$ and $N$ [Fig.~\ref{fig:dynamic}(a)]. At low $p$ the protocols are comparable. As $p$ grows the DODAG-X advantage becomes pronounced and increases with $N$: at $N=29$ the gap grows from $1.27$ measurements at $p=0.01$ to $4.69$ at $p=0.90$. Both the path length and the isolating measurement count fall for the shallow, highly branched trees produced at large $p$.

\subsection{Dynamic link failure}\label{sec:dynamic}

\begin{figure*}[t]
  \centering
  \includegraphics[width=\linewidth]{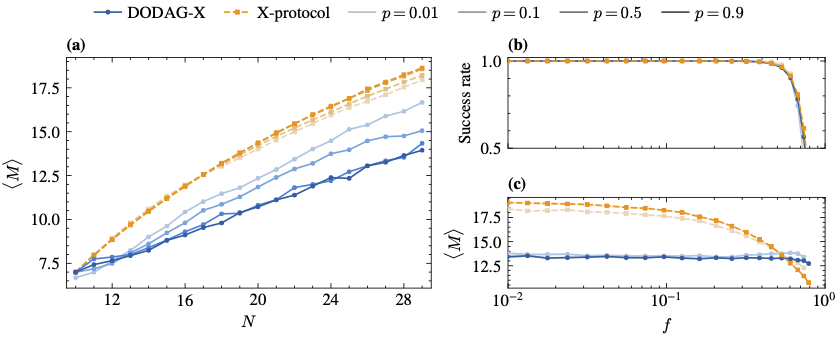}
  \caption{Measurement cost and routing success for DODAG-X (blue, solid,
  circles) and the X-protocol (orange, dashed, squares). Line lightness encodes
  the Watts--Strogatz rewiring probability $p$, lightest at $p=0.01$ and darkest
  at $p=0.9$. \textbf{(a)}~Mean measurements per request on static graphs at
  fixed mean degree $k=9$, varying $N$ ($40$ seeded instances per point). The
  X-protocol is nearly flat in $p$, while the DODAG-X advantage widens with both
  $N$ and $p$. \textbf{(b)},\textbf{(c)}~Dynamic link failure on Watts--Strogatz
  graphs ($N=30$, $k=9$, $p_{\text{rec}}=0.1$, $20$ graph instances), against the
  stationary failed-edge fraction
  $f=p_{\text{fail}}/(p_{\text{fail}}+p_{\text{rec}})$.
  \textbf{(b)}~Routing success rate with $95\%$ cluster-bootstrap confidence
  intervals resampled by graph instance. Tree connectivity implies graph
  connectivity but not the reverse, so DODAG-X is upper-bounded by full-graph
  re-search by construction; the two are statistically equivalent up to an
  empirical threshold of $f\approx0.5$ and differ by at most $7$ percentage
  points beyond it.
  \textbf{(c)}~Mean measurements per successful request, conditioned on
  requests where both protocols succeed and truncated where success rates fall
  below $20\%$. DODAG-X retains its measurement advantage across the regime where
  both protocols route reliably; in the deep-failure tail the X-protocol cost
  falls below it, because the few surviving requests are disproportionately
  short-path triplets.}
  \label{fig:dynamic}
\end{figure*}

The benchmarks above assume a static graph. We now allow physical links to fail and recover over time, using a standard discrete-time Markov model for link availability \cite{holme2012temporal, clementi2010flooding, callaway2000network}. Each physical edge fails independently with probability $p_{\text{fail}}$ per timestep and an absent edge recovers with probability $p_{\text{rec}}$, giving a stationary failed-edge fraction $f=p_{\text{fail}}/(p_{\text{fail}}+p_{\text{rec}})$. When a tree edge fails, the orphaned node reconnects through its shallowest surviving physical neighbor, repairing the DODAG on the surviving graph. The X-protocol instead re-searches the surviving graph from scratch on every request. Both protocols see identical failure histories and identical triplet (GHZ) requests ($N=30$, $k=9$, $T=5000$ timesteps, request rate $0.1$). Results are measured after a $1000$-step warm-up period during which the edge-failure process and the repaired-tree structure relax to their stationary regimes.

DODAG-X succeeds on a request only if the parties remain connected through the repaired tree, whereas the X-protocol succeeds whenever they remain connected anywhere in the surviving graph. Tree connectivity implies graph connectivity but not the reverse, so the DODAG-X success condition is strictly stronger and its reachability is upper-bounded by full re-search by construction; the experiment quantifies the resulting gap. Fig.~\ref{fig:dynamic}(b) shows the two success rates to be statistically equivalent up to $f\approx0.5$, using the two one-sided tests (TOST) procedure with a $1\%$ equivalence margin \cite{lakens2017equivalence}, with a bounded reachability gap of at most $7$ percentage points in the deep-failure tail, where both protocols already fail the majority of requests. Across the entire sweep no request succeeded for DODAG-X while failing for the X-protocol, consistent with the strict ordering.

The measurement advantage persists. On requests where both protocols succeed, DODAG-X uses ${\sim}25\%$ fewer measurements at moderate failure fractions [Fig.~\ref{fig:dynamic}(c)], comparable in magnitude to the static reduction of Sec.~\ref{sec:benchmarks}. At high $f$ the X-protocol cost falls below DODAG-X because the few surviving requests are disproportionately short-path triplets. The repair layer therefore preserves the measurement advantage of the static protocol while matching idealized full re-search on reachability, and does so without ever re-verifying global topology. The whole regime is bounded by the coherence condition $T_2\gg d\cdot\tau_c$ of Sec.~\ref{sec:discussion}, since the repair messaging must complete within the coherence window for the tree to remain routable.

\section{Discussion \& Conclusion}\label{sec:discussion}

DODAG-X routes each party to its nearest intersection node on a topology-preserving spanning tree, entangles the intersection nodes with the root, and isolates the parties by measurement. Because the tree is built once, all party-to-root paths come from a single classical computation, and the per-request path search and graph re-verification of \cite{hahn2019quantum,mannalath2023multiparty} are never repeated. The scheme is compatible with the link-loss maintenance of \cite{yang2023asynchronous}, whose repair layer matches full-graph re-search on reachability up to an empirical threshold of $f\approx0.5$ while preserving the measurement advantage (Sec.~\ref{sec:dynamic}). Since the persistent-tree connectivity condition is strictly stronger, DODAG-X trails by at most $7$ percentage points, and only where both protocols already fail the majority of requests. The operational regime can nonetheless be bounded without a decoherence model. Classical messages used by the DODAG maintenance layer to repair a broken link must propagate up and down the tree, incurring a latency of order $d \cdot \tau_c$, where $d$ is the tree depth and $\tau_c \approx L/c_{\text{f}}$ is the single-hop classical signaling time for internode distance $L$ and in-fiber signal speed $c_{\text{f}}\approx 2\times10^{8}\,\mathrm{m/s}$. The protocol is viable when the hardware coherence time satisfies $T_2 \gg d \cdot \tau_c$. For $L=10$\,km in fiber and a small-world tree of depth $d\approx5$, this gives $d\cdot\tau_c\approx250$\,$\mu$s, comfortably satisfied by trapped-ion qubits ($T_2\sim10$\,min \cite{wang2020single}), silicon spin qubits ($T_2\sim2$\,s \cite{tyryshkin2012electron}), and NV centers under dynamical decoupling ($T_2\sim1$\,s \cite{abobeih2018one}), though marginal for superconducting transmon qubits ($T_2\sim1$\,ms \cite{tuokkola2024superconducting}).

Several directions follow directly from the analysis above. The most immediate is optimizing the spanning tree itself. We fix a single minimum-eccentricity root and a BFS tree, but Sec.~\ref{sec:benchmarks} shows the measurement count is sensitive to that choice, and a construction that minimizes total neighborhood size instead of depth may do better at the cost of a more expensive one-time setup, which the amortization argument can absorb. A second direction is Step~\ref{p:5}. As specified it contracts a single non-root intersection node, which suffices for $n\le3$ (Theorem~\ref{thm:step2}) and, empirically, for party sets with at most one such node (Appendix~\ref{app:higher}). Extending it to several distinct intersection nodes would widen the single-application regime, but Appendix~\ref{app:higher} shows the difficulty is not merely procedural: for $n\ge4$ a connected output no longer determines the entanglement class, so any such extension must track the local-Clifford class as well as connectivity. A protocol-level predicate deciding, from the branch structure of a party set alone, which LC class the output falls in would settle this; we do not provide one.

When one application is insufficient, a concatenated construction, i.e., multiple qubits per DODAG node with the outputs of separate runs merged locally, is the natural fallback, and costing it against the direct approach is open. Relatedly, the protocol treats requests one at a time, whereas a realistic network serves many concurrently; because Step~\ref{p:2} may re-root per request, scheduling several requests against one tree, and quantifying the contention when their paths overlap, is a concrete next step. Finally, our cost accounting is measurement-based, following \cite{hahn2019quantum,mannalath2023multiparty}. A full hardware cost model would also count the entangling operations used to prepare and maintain the resource state and propagate a per-operation error rate through to output fidelity, which requires a decoherence model we have not attempted. Since DODAG-X maintains only $N-1$ links against the physical graph's $|E|$, we expect this accounting to favor it further, but we do not claim it here.

Beyond entanglement, the same framework may route other quantum resources, for instance non-stabilizerness, or ``magic'' \cite{Bravyi:2004isx,Oliviero:2022euv,Oliviero:2022bqm}, which is tied to entanglement structure \cite{Bao_2022,Cao:2024nrx}. Graph states also admit a group-theoretic description \cite{hein2006entanglement,Munizzi:2023ihc,Keeler:2022ajf,Keeler:2023xcx,Latour:2022gsf} through which global network constraints \cite{Keeler:2023shl} and entanglement rates \cite{Couch:2019zni} could be analyzed.

\section*{Acknowledgments}
We thank Dr.\ Ze-Xun Lin for helpful conversations. We acknowledge NSF Awards 2137828, 2246394, and 2107265, and P.N.\ acknowledges Grant No.\ GBMF8048 from the Gordon and Betty Moore Foundation. Code and data are available upon request.

\bibliographystyle{JHEP}
\bibliography{references}

\newpage

\appendix

\section{Path-finding complexity}\label{app:complexity}

The eccentricity of $v$ is $\mathrm{Ecc}(G,v)=\max_{u}d(v,u)$. The root minimizes it, after which path-finding depends only on tree depth. For an $m\times\ell$ grid a root of minimum eccentricity sits at the lattice center, where the depth is $\lfloor (m-1)/2\rfloor+\lfloor(\ell-1)/2\rfloor$. The maximal asymmetry ($m$ or $\ell$ equal to $1$) gives depth $\sim N/2$ and $\mathcal{O}(N)$. For $m=\ell$ the depth is $\sim m=\sqrt N$, giving $\mathcal{O}(\sqrt N)$, and intermediate ratios interpolate. For Watts--Strogatz small-world graphs the average internode distance scales as $\log N$ \cite{watts1998collective}, so the average depth is $\mathcal{O}(\log N)$.

\section{Proofs of connectivity}\label{app:proofs}

Throughout this appendix, $N_v^{(m)}$ denotes the neighborhood of $v$ after $m$
measurements and $N_v^{(u)}$ the neighborhood of $v$ after measuring the specific
node $u$ (Sec.~\ref{sec:graphs}). For a tree path $P=(v_1,\dots,v_m)$,
\begin{equation}\label{eq:tree}
N_{v_t}\cap N_{v_{t+2}}=\{v_{t+1}\},\quad \forall t\in\{1,\dots,m-2\},
\end{equation}
and, since a tree contains no cycles, $v_1$ is adjacent to no path vertex other
than $v_2$. Lemma~\ref{lem:evolution} extends the neighborhood analysis of
\cite{hahn2019quantum,mannalath2023multiparty} to this setting.

\begin{lemma}[Neighborhood evolution]\label{lem:evolution}
Let $P=(v_1,\dots,v_m)$ be a path in a tree whose interior nodes
$v_2,\dots,v_{m-1}$ are sequentially $X$-measured as in Step~\ref{p:4}. Write
$A\Delta B\equiv(A\cup B)\setminus(A\cap B)$; since $\Delta$ is associative and
commutative, the iterated symmetric difference $\bigDelta_{t\in S}N^{(0)}_{v_t}$
over an index set $S$ is unambiguous. The endpoint neighborhoods after the $m-2$
measurements are
\begin{equation}\label{eq:Nv1}
N^{(m-2)}_{v_1}=
\begin{cases}
\bigDelta_{\substack{t=1\\ t\ \mathrm{odd}}}^{m-1} N^{(0)}_{v_t}, & m\ \text{even},\\[6pt]
\Big(\bigDelta_{\substack{t=2\\ t\ \mathrm{even}}}^{m-1} N^{(0)}_{v_t}\Big)\setminus\{v_1\}, & m\ \text{odd},
\end{cases}
\end{equation}
\begin{equation}\label{eq:Nvm}
N^{(m-2)}_{v_m}=
\begin{cases}
\bigDelta_{\substack{t=2\\ t\ \mathrm{even}}}^{m} N^{(0)}_{v_t}, & m\ \text{even},\\[6pt]
\{v_1\}\cup\Big(\bigDelta_{\substack{t=1\\ t\ \mathrm{odd}}}^{m} N^{(0)}_{v_t}\Big), & m\ \text{odd}.
\end{cases}
\end{equation}
In each case the index runs over the path vertices of one parity, ending at
$v_{m-1}$ in Eq.~\eqref{eq:Nv1} and at $v_m$ in Eq.~\eqref{eq:Nvm}. Since $v_1$
lies in $N^{(0)}_{v_2}$ but in no other $N^{(0)}_{v_t}$ of the chain, the
deletion $\setminus\{v_1\}$ in the odd case of Eq.~\eqref{eq:Nv1} may
equivalently be applied to $N^{(0)}_{v_2}$ alone.
\end{lemma}
\noindent\emph{Proof.} The single-measurement update underlying
Eqs.~\eqref{eq:Nv1}--\eqref{eq:Nvm}, namely
$N'_{v_1}=N_{v_t}\setminus\{v_1\}$ and
$N'_{v_{t+1}}=\{v_1\}\cup\big(N_{v_{t+1}}\Delta N_{v_1}\big)$ for
$X[v_t,v_1]$, is derived in Eqs.~(12)--(14) of the Supplementary Information
of \cite{hahn2019quantum}, stated there for a shortest path with the
complementation basis held fixed at $v_1$ throughout; see also
\cite{mannalath2023multiparty}. On a tree every path is the unique, hence
shortest, path between its endpoints, so that derivation applies.
Eqs.~\eqref{eq:Nv1}--\eqref{eq:Nvm} follow by iterating it along $P$: the
alternating substitution of the update into itself produces the
symmetric-difference chain over path vertices of one parity, and the
preconditions of the update persist at every step, since on a tree
Eq.~\eqref{eq:tree} excludes any adjacency among $v_1$ and the path vertices 
beyond $v_2$. Each measurement in the sequence is in particular well defined:
$v_1\in N^{(0)}_{v_2}$ initially, and applying Eq.~\eqref{eq:Nvm} to the
prefix $(v_1,\dots,v_{t+1})$, whose remainder is untouched by the earlier
measurements, shows that $v_1\in N_{v_{t+1}}$ once $v_t$ has been measured,
so the basis $v_1$ is adjacent to each interior vertex at the moment it is
measured. $\square$

\begin{lemma}[Entangled pairs]\label{lem:pairs}
On a tree, $X$-measuring the interior of the path $P=(v_1,\dots,v_m)$ between
two nodes $v_1$ and $v_m$ entangles the pair in at most $m-2$ measurements.
\end{lemma}
\noindent\emph{Proof.} It suffices to show $v_1\in N^{(m-2)}_{v_m}$ after the
measurements. For $m$ odd this is immediate from Eq.~\eqref{eq:Nvm}, whose odd
case contains $\{v_1\}$ explicitly. For $m$ even, $v_1\in N^{(0)}_{v_2}$ by
construction of the path, while $v_1\notin N^{(0)}_{v_s}$ for every path vertex
$v_s$ with $s\neq2$ by Eq.~\eqref{eq:tree}; hence $v_1$ enters the
symmetric-difference chain of Eq.~\eqref{eq:Nvm} exactly once and survives it.
$\square$

\begin{theorem}\label{thm:step1}
After entangling each party $i$ with its nearest intersection node $a_{i,j}$, each party is connected to an intersection node or is one itself.
\end{theorem}
\noindent\emph{Proof sketch.} We route first the parties that are not
themselves intersection nodes, as specified in Step~\ref{p:4}. For such a
party, the sequential $X$-measurements along $P_{i,a_{i,j}}$ (excluding
endpoints) connect $i$ to $a_{i,j}$ (Lemma~\ref{lem:pairs}), reproducing the
connectivity result of the X-protocol \cite{hahn2019quantum}.

The remaining case is a party that is itself an intersection node, $i=a_{i,j}$
for some other party $j$: after Step~\ref{p:4} the neighborhood $N_i$ contains
$j$, and must be tracked through the measurements along the path from $i$ to
its own nearest intersection node $a$. Writing that path as
$P=(v_1{=}i,\dots,v_m{=}a)$ and combining Lemma~\ref{lem:evolution} with
Eq.~\eqref{eq:tree}, the set $N^{(0)}_{v_1=i}\setminus\{v_2\}$ is contained in
$N^{(m-2)}_{v_m=a}$ when $m$ is odd, and in $N^{(m-2)}_{v_1=i}$ when $m$ is
even. The party $j$ that had $i$ as its intersection node therefore remains
connected, after the measurements, to either $i$ or $a$; both are
intersection nodes, so every party ends connected to an intersection node.
$\square$

\begin{corollary}
For three parties, if an intersection node is also a party it is entangled with the root after Step~\ref{p:4}.
\end{corollary}
\noindent For $n$ parties there are at most $n-1$ distinct intersection nodes in total, so with $n=3$ there are at most two, one of which must be the root after the re-rooting of Step~\ref{p:2}; hence a party-intersection node has the root as its nearest intersection. $\square$

\begin{theorem}\label{thm:step2}
For $n\le3$, DODAG-X always yields the parties entangled with each other and with no other node.
\end{theorem}
\noindent\emph{Proof.} Cases 5(a),(b) need no extra Step-\ref{p:5}
measurement (Theorem~\ref{thm:step1}). For case 5(c) we track the alternating
$X$-measurements along $P_{a_{i,j},r}=\{v_1{=}a_{i,j},\dots,v_m{=}r\}$,
writing $N^{(0)}$ throughout this case for neighborhoods at the start of
Step~\ref{p:5}, that is, after the measurements of Step~\ref{p:4}. Each
measurement is well defined, since the basis $v_{s-1}$ of $X[v_s,v_{s-1}]$ is
an unmeasured path vertex joined to $v_s$ by the tree edge
$\{v_{s-1},v_s\}$, which no earlier measurement of the chain toggles, and in
the even case $i\in N_{v_{m-1}}$ because $i\in N_{a_{i,j}}$ at the start of
Step~\ref{p:5} and the neighborhood of $a_{i,j}$ propagates to the
odd-index vertices along the recursion below. Each display is an instance of
the update rule of Lemma~\ref{lem:evolution}, whose preconditions hold at
every step by Eq.~\eqref{eq:tree}. After
$X[v_2,v_1]$,
\begin{align}
N^{(v_2)}_{v_1}&=N^{(0)}_{v_2}\setminus\{v_1\}, \nonumber\\
N^{(v_2)}_{v_3}&=\{v_1\}\cup\big(N^{(0)}_{v_3}\Delta N^{(0)}_{v_1}\big),
\end{align}
and after the next measurement, $X[v_4,v_3]$,
\begin{align}
N^{(v_4)}_{v_3}&=N^{(0)}_{v_4}\setminus\{v_3\}, \nonumber\\
N^{(v_4)}_{v_5}&=\{v_3\}\cup\big(N^{(0)}_{v_5}\Delta N^{(v_2)}_{v_3}\big),
\end{align}
with the same pattern continuing down the path. For $m$ odd the last measured
node is $v_{m-1}$, giving
\begin{align}
N^{(v_{m-1})}_{v_{m-2}}&=N^{(0)}_{v_{m-1}}\setminus\{v_{m-2}\}, \nonumber\\
N^{(v_{m-1})}_{v_m}&=\{v_{m-2}\}\cup\big(N^{(0)}_{v_m}\Delta N^{(v_{m-3})}_{v_{m-2}}\big),
\end{align}
so both $a_{i,j}$ and $N^{(0)}_{a_{i,j}}$ end in the final neighborhood of the
root, while $N^{(0)}_{r}\setminus\{v_{m-1}\}$ remains connected to it. For $m$
even the same recursion runs to $v_{m-2}$, after which the final $X[v_{m-1},i]$
of Step~\ref{p:5} yields
\begin{align}
N^{(v_{m-1})}_{i}&=N^{(v_{m-2})}_{v_{m-1}}\setminus\{i\}, \nonumber\\
N^{(v_{m-1})}_{v_m}&=\{i\}\cup\big(N^{(0)}_{v_m}\Delta N^{(v_{m-2})}_{i}\big).
\end{align}
The set $N^{(v_{m-2})}_{v_{m-1}}\setminus\{i\}$, which contains party $j$,
becomes part of $N_i$, connecting $j$ to $i$, while $i$ is connected directly to
$r$.

In both parities the elements carried along the chain survive it: an element
of $N_{a_{i,j}}$ enters the nested symmetric differences exactly once, since a
second occurrence would require a further path vertex adjacent to it and hence
a cycle, which a tree forbids. The parties therefore reach Step~\ref{p:6} with
the induced subgraph on $\Parties\cup\{r\}$ connected. In case 5(a) each party
is adjacent to $r$ by Theorem~\ref{thm:step1}; in case 5(b) the
party-intersection node is adjacent to $r$ and carries the party that met it;
in case 5(c) the recursions above leave $a_{i,j}$ and its neighbors, among them
$i$ and $j$, in $N_r$ for $m$ odd, and leave $i$ adjacent to $r$ with $j\in
N_i$ for $m$ even, while the third party is adjacent to $r$ from
Step~\ref{p:4}. Up to relabeling of the parties, the induced subgraph on
$\Parties\cup\{r\}$ at this point is therefore either the star centered at $r$,
with edge set $\{i,r\},\{j,r\},\{k,r\}$ [cases 5(a), and 5(b),(c) with $m$
odd], or the graph with edge set $\{i,r\},\{k,r\},\{i,j\}$ [cases 5(b),(c) with
$m$ even].

If $r\in\Parties$, Step~\ref{p:6} is skipped and the parties already induce a
connected subgraph. Otherwise it performs
$X[r,i_1]=\tau_{i_1}Z_r\tau_r\tau_{i_1}$ for a party $i_1\in N_r$; in the even
subcases the party $i$ used in Step~\ref{p:5} qualifies. An edge between two
tracked vertices is toggled only when both of its endpoints lie in the
neighborhood of the vertex being complemented, so the induced subgraph on
$\Parties\cup\{r\}$ evolves autonomously under the three complementations of
$X[r,i_1]$: vertices outside $\Parties\cup\{r\}$ cannot alter party--party
adjacency, and the effect of Step~\ref{p:6} is fixed by a graph on at most four
vertices. It suffices to apply $\tau_{i_1}Z_r\tau_r\tau_{i_1}$ to the two
configurations above. In the star, $\tau_{i_1}$ finds only $r$ among the
tracked vertices and toggles nothing; $\tau_r$ makes the three parties pairwise
adjacent; $Z_r$ deletes $r$; and the final $\tau_{i_1}$, whose tracked
neighborhood is now the other two parties, toggles that pair off again, leaving
the star centered at $i_1$. In the second configuration the admissible choices
are $i_1=i$ and $i_1=k$, giving the paths $i$--$k$--$j$ and $j$--$i$--$k$
respectively. In every reachable configuration the parties induce a connected
graph.

Step~\ref{p:7} $Z$-measures every neighbor of a party that is not itself a
party. Vertex deletion removes no edge between two surviving vertices
[Eq.~\eqref{eq:delete}], so no party--party edge is affected, while by
construction no party retains a neighbor outside $\Parties$. The output is a
connected graph on the three parties, disconnected from the rest of the
network, hence $P_3$ or $K_3$; both are LC-equivalent to GHZ$_3$
\cite{hein2004multiparty}. $\square$

\begin{lemma}[Branch locality]\label{lem:branch}
Let $r$ be the root and let $B$ be the vertex set of the subtree rooted at one
child of $r$, so that initially $N_u\subseteq B\cup\{r\}$ for every $u\in B$.
Any sequence of measurements $X[v,w]$ or $Z[v]$ with $v,w\in B$ preserves this
property, creates or removes no edge lacking an endpoint in $B$, and creates no
edge with an endpoint outside $B\cup\{r\}$.
\end{lemma}
\noindent\emph{Proof.} A $Z$-measurement only deletes a vertex of $B$. An
$X$-measurement $X[v,w]=\tau_w Z_v\tau_v\tau_w$ [Eq.~\eqref{eq:x}] is a
composition of local complementations at vertices of $B$ and one such deletion.
A local complementation at $u\in B$ toggles only pairs within
$N_u\subseteq B\cup\{r\}$; every toggled pair therefore lies in $B\cup\{r\}$
and has at least one endpoint in $B$. The invariant is maintained inductively,
and in particular no edge between two distinct branches, or between $r$ and a
vertex outside $B$, is ever created or removed. $\square$

\begin{theorem}[Single-shot GHZ$_n$]\label{thm:ghzn}
Let $\Parties=\{i_1,\dots,i_n\}$, $n\ge2$, and suppose all pairwise
intersection nodes coincide: $a_{i_s,i_t}=a$ for all $s\neq t$. Then DODAG-X
yields the parties in the star $K_{1,n-1}$, locally equivalent to GHZ$_n$,
isolated from the rest of the network. The star is centered at $a$ if
$a\in\Parties$, and otherwise at the party used for local complementation in
Step~\ref{p:6}.
\end{theorem}
\noindent\emph{Proof.} By hypothesis $a$ lies on every $P_{i,r}$, while any
vertex strictly below $a$ lies in a single branch of $a$ and hence on at most
one party's path; $a$ is therefore the common vertex farthest from $r$, and
the re-rooting of Step~\ref{p:2} sets the root to $a$. Without loss of
generality $r=a$. The least element of each pairwise path intersection is then
$r$ itself, so the party-to-root paths meet only at $r$: each party
$i_s\neq a$ lies in a distinct branch $B_s$ of $r$, and vertices in distinct
branches are non-adjacent initially.

Step~\ref{p:4} entangles each party $i_s\neq a$ with $r$ by $X$-measuring the
interior of $P_{i_s,r}$, all at vertices of $B_s$ with basis $i_s\in B_s$. By
Lemma~\ref{lem:branch} the operations for branch $B_s$ alter no edge incident
to a vertex of another branch and add to $N_r$ only vertices of $B_s$; the
internal structure of every other branch, and the path property
Eq.~\eqref{eq:tree} along its party-to-root path, are therefore intact when its
turn comes. The vertices that earlier branches have added to $N_r$ are
invisible to the later ones, since the chain of Lemma~\ref{lem:evolution} for
branch $B_s$ involves only the neighborhoods of vertices of $B_s$ and the
membership of $i_s$ in them. Lemma~\ref{lem:pairs} thus applies branch by
branch: after
Step~\ref{p:4} every party other than $r$ is adjacent to $r$, while these
parties remain pairwise non-adjacent, since no inter-branch edge is ever
created.

All intersection nodes equal $r$, so Step~\ref{p:5} is case (a) and requires
no measurement. Two cases remain.

If $r\in\Parties$, Step~\ref{p:6} is skipped, and the isolating
$Z$-measurements of Step~\ref{p:7} delete only non-party vertices: the parties
survive with $r$ adjacent to each of the others and the others pairwise
non-adjacent, the star centered at $r=a$.

If $r\notin\Parties$, Step~\ref{p:6} performs
$X[r,i_1]=\tau_{i_1}Z_r\tau_r\tau_{i_1}$ for a party $i_1\in N_r$. The first
$\tau_{i_1}$ toggles pairs within $N_{i_1}\subseteq B_1\cup\{r\}$
(Lemma~\ref{lem:branch}), touching no edge incident to another party and, since
local complementation never alters edges at its own vertex, leaving every edge
$\{i_s,r\}$ intact. The complementation $\tau_r$ then toggles all pairs within
$N_r\supseteq\Parties$: every pair of parties, previously non-adjacent, becomes
adjacent. $Z_r$ deletes $r$. The final $\tau_{i_1}$ toggles pairs within the
new $N_{i_1}$, which contains every other party; each pair $\{i_s,i_t\}$ with
$s,t\ge2$ is toggled off again, while the edges at $i_1$ itself are untouched.
The parties therefore form the star centered at $i_1$, and Step~\ref{p:7}
removes every remaining non-party neighbor without affecting party--party
edges. $\square$

\noindent For $n\le3$ the hypothesis reduces to case 5(a), consistent with
Theorem~\ref{thm:step2}; for $n\ge4$ it is sufficient but not necessary, and
Appendix~\ref{app:higher} examines what happens outside it.

\section{Measurement counts}\label{app:counts}

\begin{theorem}\label{thm:measurement}
DODAG-X and the X-protocol require the same number of measurements for $n\le3$ on the same tree.
\end{theorem}
\noindent\emph{Path measurements.} For the X-protocol, $M=|P_{ij}\cup(P_{jk}\setminus(P_{ij}\cap P_{jk}))|-n=|P_{ij}\cup P_{jk}|-n$. For DODAG-X, $M'=|P_{i,r}\cup P_{j,r}\cup P_{k,r}|-n$. On a tree, $P_{i,r}\setminus P_{j,r}=P_{i,a_{i,j}}\setminus\{a_{i,j}\}$, so $P_{ij}=P_{i,a_1}\cup P_{j,a_1}$ and $P_{jk}=P_{j,a_2}\cup P_{k,a_2}$ with $a_1=a_{i,j}$, $a_2=a_{j,k}$. Since the network is a tree, $a_1$ and $a_2$ both lie on $P_{j,r}$, so one is an ancestor of the other; without loss of generality take $a_1$ to lie on $P_{j,a_2}$. Writing $P_{j,a_2}=P_{j,a_1}\cup P_{a_1,a_2}$ gives $P_{ij}\cup P_{jk}=P_{i,a_2}\cup P_{j,a_2}\cup P_{k,a_2}$. Since three parties admit at most two distinct intersection nodes and the root lies in every pairwise path intersection, $a_2$ may be relabeled as $r$ without loss of generality (cf.\ the re-rooting of Step~\ref{p:2}), giving $P_{ij}\cup P_{jk}=P_{i,r}\cup P_{j,r}\cup P_{k,r}$, hence $M=M'$.

\emph{Isolation measurements.} The isolating $Z$-measurements equal the total path neighborhood minus the measured path nodes,
\begin{equation}
|N_T|-|P_{i,r}\cup P_{j,r}\cup P_{k,r}|=|N_T|-|P_{ij}\cup P_{jk}|,
\end{equation}
with $N_T=\bigcup_{v\in P_{i,r}\cup P_{j,r}\cup P_{k,r}}N_v^{(0)}$ \cite{mannalath2023multiparty}. The two counts coincide on a tree. $\square$

\emph{Correspondence with the protocol.} The count $M'$ charges every non-party
vertex of the path union as consumed, which matches the protocol exactly.
Steps~\ref{p:4} and~\ref{p:6} $X$-measure the path interiors and the root; the
only path vertices left un-$X$-measured are the odd-index vertices of
$P_{a_{i,j},r}$ in case 5(c), including $a_{i,j}$ itself when
$|P_{a_{i,j},r}|$ is odd. By the recursion in the proof of
Theorem~\ref{thm:step2}, each such vertex $v_{s-1}$ enters the surviving chain
exactly once, through the $\{v_{s-1}\}$ term created when $v_s$ is measured,
and by acyclicity is never cancelled from the nested symmetric differences. It
therefore ends Step~\ref{p:5} adjacent to the root or to a party, is adjacent
to a party at the latest after the root measurement of Step~\ref{p:6}, and is
consumed by the isolating $Z$-measurements of Step~\ref{p:7}. Every vertex of
$N_T\setminus\Parties$ is thus measured exactly once, and the two totals agree.

\section{Higher-party examples}\label{app:higher}

\begin{figure*}[tp]
\centering
\includegraphics[width=\textwidth]{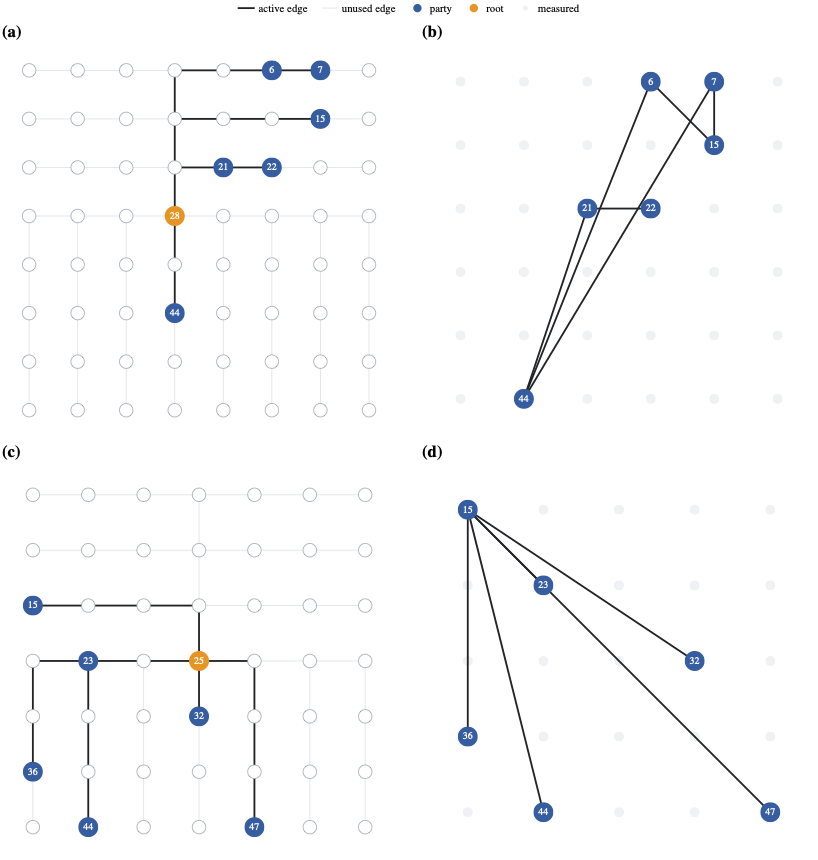}
\caption{Six-party outputs on the benchmark grids; conventions as in
Fig.~\ref{fig:example}, but drawn in lattice coordinates.
\textbf{(a)},\textbf{(b)}~Six parties on an $8\times8$ grid whose
party-to-root paths define four distinct non-root intersection nodes, outside
the hypothesis of Theorem~\ref{thm:ghzn}. \textbf{(a)}~the DODAG with the
parties (blue) and root $v_{28}$ (orange) marked; \textbf{(b)}~the isolated
output, connected, hence genuinely six-party entangled, but LC-equivalent to
the six-qubit linear cluster state rather than GHZ$_6$.
\textbf{(c)},\textbf{(d)}~Six parties on a $7\times7$ grid with a single
non-root intersection node, $a_{36,44}=v_{23}$, which is itself a party, so
that Step~\ref{p:5} applies as case (b); every other pairwise intersection
node is the root $v_{25}$. This configuration also lies outside the hypothesis
of Theorem~\ref{thm:ghzn}, which requires all pairwise intersection nodes to
coincide. \textbf{(c)}~the DODAG; \textbf{(d)}~the isolated output, the star
$K_{1,5}$ centered at $v_{15}$, the party used for local complementation in
Step~\ref{p:6}, locally equivalent to GHZ$_6$, in agreement with the
theorem's conclusion if not its hypothesis.}
\label{fig:higher_examples}
\end{figure*}

Outside the hypothesis of Theorem~\ref{thm:ghzn}, the class of the output state
is no longer fixed by its connectivity. The examples of
Fig.~\ref{fig:higher_examples}, on the same grid topologies as
Sec.~\ref{sec:benchmarks}, bound the region in which the conclusion of the
theorem still holds.

Both branches of Theorem~\ref{thm:ghzn} hold directly on the $7\times7$ DODAG
of Fig.~\ref{fig:higher}, checked on party sets constructed to satisfy its
hypothesis. With $a\notin\Parties$, three four-party sets whose pairwise
intersection nodes all coincide at $a=v_{25}$ return the star centered at the
Step-\ref{p:6} party. With $a\in\Parties$, Step~\ref{p:6} is skipped and the
star is centered at $a$ itself, both for $a=v_{25}$ equal to the root and for
$a=v_{11}$ distinct from it.

The instances shown in the figures lie just outside the hypothesis and still
reach its conclusion. Six parties on a $7\times7$ grid with a single non-root
intersection node, itself a party [Fig.~\ref{fig:higher_examples}(c,d)],
produce the star $K_{1,5}$ centered at the Step-\ref{p:6} party, GHZ$_6$ in one
application; the four-party request of Fig.~\ref{fig:higher}, whose single
non-root intersection node is not a party, produces $K_{1,3}$ centered at the
Step-\ref{p:6} party. Across the instances we tested, the star is obtained
exactly when at most one non-root intersection node is present, but we have no
proof that this weaker condition is sufficient in general.

Further out, six parties on an $8\times8$ grid whose paths define four distinct
non-root intersection nodes [Fig.~\ref{fig:higher_examples}(a,b)] still return
an isolated, connected output, hence a genuinely six-party entangled state
\cite{hein2006entanglement}, but exhaustive enumeration of its
local-complementation orbit ($82$ graphs, containing no star) identifies it as
the six-qubit linear cluster state, an LC class distinct from GHZ$_6$
\cite{hein2004multiparty}. Connectivity of the output therefore does not imply
GHZ$_n$ for $n\ge4$; which class is produced depends on the branch structure of
the party set.

A third configuration fails outright: six parties whose paths define two
distinct non-root intersection nodes. Step~\ref{p:5} as specified contracts one
of them, and the branch reaching the root only through the other is severed by
the isolating measurements of Step~\ref{p:7}, leaving GHZ$_4$ on four of the
parties with the remaining two disentangled. Detecting this outcome requires a
connectivity check on the output, which the protocol as specified does not
include; extending Step~\ref{p:5} to several intersection nodes remains open.

\end{document}